\documentclass[usenatbib]{mn2e}

\usepackage{epsfig}
\usepackage{amsmath}
\usepackage{natbib}
\usepackage{times}

\newcommand{\ms}{h^{-1} {\rm M_{\odot}}}

\def\gsim { \lower .75ex \hbox{$\sim$} \llap{\raise .27ex \hbox{$>$}}}
\def\lsim { \lower .75ex \hbox{$\sim$} \llap{\raise .27ex \hbox{$<$}}}
\title
{Assembly Bias of Dwarf-sized Dark Matter Haloes}
\author[Ran Li et al.]
       {Ran Li$^{1}$ \thanks{Email:ranli@bao.ac.cn}, 
        Liang Gao$^{1,2}$,
        Lizhi Xie$^1$,
        Qi Guo$^{1,2}$
\\
$^1$The Partner Group of Max Planck Institute for Astrophysics,
National Astronomical Observatories, Chinese Academy of Sciences, Beijing, 100012, China\\
$^2$Institute of Computational Cosmology, Department of Physics,
University of Durham, Science Laboratories, South Road, Durham DH1
3LE \\
}

\begin{document}
\maketitle
\begin{abstract}
Previous studies indicate that assembly bias effects are stronger
for lower mass dark matter haloes. Here we make use of high resolution
re-simulations of rich clusters and their surroundings from the
Phoenix Project and a large volume cosmological simulation, the
Millennium-II run, to quantify assembly bias effects on dwarf-sized 
dark matter haloes. We find that, in the regions around massive clusters,
 dwarf-sized haloes ( $[10^9,10^{11}]\ms$) form earlier ($\Delta z \sim 2$ in
redshift) and possess larger $V_{\rm max}$ ($\sim20\%$)  than the
field galaxies.  We find that this environmental dependence is
  largely caused by tidal interactions between the ejected haloes and
  their former hosts, while other large scale effects are less
  important. Finally we assess  the effects of assembly bias on dwarf
galaxy formation with a sophisticated semi-analytical galaxy formation
model.  We find that the dwarf galaxies near massive clusters tend to
be redder ($\Delta(u-r) = 0.5$) and have three times as much stellar
mass compared to the field galaxies with the same halo mass. These
features should be seen with  observational data.
\end{abstract}


\section{Introduction}
Halo assembly bias describes the phenomenon that the clustering of 
dark matter halo depends not only on their halo mass but also on the properties, such
as formation time, concentration parameter, spin and substructure fraction
 \citep[e.g.][]{Gao2005,Harker2006,Wechsler2006,Gao2007,Jing2007,Li2008}. 
For a given halo mass, earlier-forming dark matter haloes are more strongly clustered.
The difference in clustering with formation time becomes larger as halo mass
 decreases \citep[e.g.][]{Gao2005}. These results contradict the standard
excursion set theory of structure formation and the theory behind the so-called
halo occupation distribution model, in which the galaxy content of a halo
is assumed to be statistically independent of its large scale
environment  \citep[e.g.][]{Jing1998, Peacock2000, Yang2003}.

A number of theoretical studies have been carried out to
understand this phenomenon  \citep[e.g.][]{Wang2007, Sandvik2007,
  Desjacques2008, Wang2009, Dalal2008, Hahn2007, Hahn2009,
  Ludlow2009,Tinker2011,Lacerna2011,Lacerna2012}. For example,
\citet{Wang2007} found that old low mass haloes tend to reside next to
massive haloes. The larger velocity dispersion and local tidal field
around clusters may suppress the halo growth of the small
haloes. Furthermore,  numerical simulations show that a fraction of
individual haloes with highly eccentric orbits around more massive
systems are former members of the system that have been expelled. 
\citep[e.g.][]{Balogh2000,Benson2005,Wang2009,Bahe2012,Sales2007}. These
haloes can lose a substantial fraction of mass, during their passage
through the larger system, which naturally induces a form of assembly
bias.  An similar effect, which is often referred to as
"pre-processing" (hereafter PREP), comes from interactions between low
mass haloes and infalling groups in the regions around
clusters\citep{Berrier2009,McGee2009}. On the observational side,
a number of studies explore the assembly bias. Results are
controversial,  some studies claim its existence \citep[e.g.][]{Yang2006,
  Wang2008, Cooper2010, Tinker2012, Kauffmann2013, Wang2013,
  Wetzel2013}, while others claim it is
non-significant\citep[e.g.][]{Blanton2007,Tinker2008,Skibba2009}.

In this work, we explore  assembly bias effects on the properties of
dwarf-sized dark matter haloes, taking advantage of extremely high
resolution re-simulations of rich clusters and their surroundings from
the Phoenix Project and the cosmological volume Millennium-II
simulation. These simulations resolve very low mass dark matter haloes
for which assembly bias effects are expected to be strong. Furthermore, we
distinguish different origins of  assembly bias according to the
assembly history of haloes.  In addition, we assess the consequences of assembly
bias on dwarf galaxy formation using a sophisticated semi-analytical
galaxy formation model(hereafter SAM).

The structure of the paper is as follows. In section~\ref{sec:method}, we
describe briefly the numerical simulations, halo samples and galaxy
formation model used in this study. In section~\ref{sec:haloresult}, we
present results of  assembly bias effects on various properties of
dwarf-sized dark matter haloes. In section~\ref{sec:galaxyresult}, we
explore how assembly bias affects the dwarf galaxy properties. 
Finally we summarize and discuss our results in section~\ref{sec:sum}.

\section{Method}
\label{sec:method}
\subsection{Numerical simulations}

Numerical simulations used in this work comprise high resolution
re-simulations of 9 individual rich clusters and their surroundings from the 
Phoenix Project \citep{Gao2012} and a high resolution cosmological
simulation -- the Millennium-II  \citep[MSII,][]{BK2009}. The dwarf-sized
haloes near the Phoenix cluster regions  are used as our high
density environment halo sample, while those in the MS-II 
represent the haloes in the field.

Each Phoenix cluster has been simulated at different
resolutions in order to  facilitate numerical convergence studies. Here,
we use the simulations with level 2 resolution which contains about
$10^8$ particles inside the virial radius of each cluster. The
least massive dwarf-sized haloes considered in this work ($10^9\ms$)
are resolved with more than 200 particles. The MS-II run evolved
$2160^3$ particles of mass $m_p=6.9 \times 10^6$ in a periodic box of
$100h^{-1}{\rm Mpc}$ on a side. Hence the mass resolutions of the
MS-II and the Phoenix Project are comparable. All these simulations adopt
identical cosmological parameters from a combination of
2dFGRS \citep{Colless2001} and first-year WMAP data \citep{spergel2003},
$\Omega_m$= 0.25, $\Omega_b$=0.045, $\Omega_{\Lambda}=0.75$, $n_s=1$,
$\sigma_8=0.9$, and  $H_0=73{\rm km s^{-1}   Mpc^{-1}}$.  
These cosmological parameters deviate slightly from the latest CMB
results\citep{Wmap2012,Planck2013}. The small offset is of no
consequence for the topic addressed in this paper because the detailed
structure of dark matter haloes depends only weakly on the cosmological
parameters\citep[e.g.][]{Duffy2008}.

Dark Matter haloes in our simulations are identified with standard
friends-of-friends(FOF) group algorithm with a linking length
0.2 times the mean inter-particle separation \citep{Davis1985}. Based
on group catalogue, we further identify self-bound substructures
within FOF haloes using SUBFIND \citep{Springel2001b} and construct
merger trees tracing subhaloes  between
snapshots \citep{Springel2005,BK2009}.  For a halo at z=0, its main
progenitor at $z=z'$ is defined as the subhalo which contains the
largest fraction of its particles.

The assembly history of small dark matter haloes in massive rich cluster regions
is complicated.  We divide the Phoenix haloes into three separated samples 
according to the different formation path: 1) A clean halo sample
(hereafter CLEAN) comprising dark matter haloes which have never be
accreted as a subhalo of a more massive FOF group at any redshift; 2)
An ejected halo sample (hereafter EJECT) comprising dark matter haloes
which were  identified at least once as a subhalo of the main progenitor of
one of the Phoenix cluster haloes. 3) A pre-processed halo sample (hereafter PREP)
comprising dark matter haloes which were identified at least once as a subhalo of a
FOF group other than the main progenitor of one of the Phoenix cluster haloes.

Since the Phoenix project  is a set of ''zoom-in'' re-simulations, only a
very small fraction of whole volume is filled with high resolution
particles, and the rest of the density field is  sampled with low 
resolution particles. For such ''zoom-in'' simulations, low resolution
particles often mix with high resolution ones  at the boundary of the 
high density regions.  It is important to identify the region where halo
samples are free from contamination by low resolution
particles. We find that, for all
simulations, there is no contamination for dark haloes within
$3r_{200}$ of the dominant clusters. Here $r_{200}$ is defined as
the radius at which the enclosed density is $200$ times of the
critical density of the universe.

\begin{figure*}
\hspace{0.13cm}\resizebox{16cm}{!}{\includegraphics{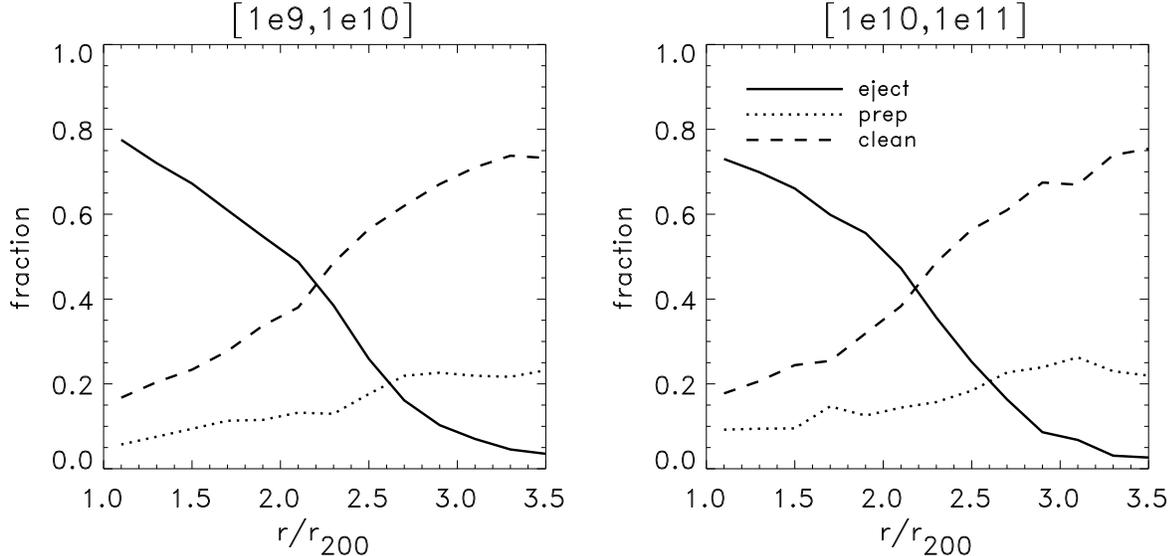}}
\caption{Fraction of EJECT , PREP and CLEAN haloes as a function of cluster-centric
 distance r, expressed in units of viral radius
$r_{200}$. Solid lines correspond to the EJECT sample, dotted lines are
for the PREP sample and dashed lines are for the CLEAN sample. Left panel
shows result for $[10^{9},10^{10}]\ms$, and right panel for
$[10^{10},10^{11}]\ms$.} \label{fig:3type} 
\end{figure*}

In Fig.~\ref{fig:3type}, we plot the fraction of these three halo
samples in the Phoenix simulations as a function of scaled radius
$r/r_{200}$ for two halo mass ranges  $[10^9, 10^{10}]
\ms$ and $[10^{10}, 10^{11}]\ms$.  As can be seen clearly from the
left-hand panel, EJECT haloes dominate the whole population 
close to the virial radius, where about $80$ percent of  dark matter
haloes have once been a member of the Phoenix clusters. The fraction of
EJECT haloes drops rapidly with  increasing radius and these haloes no longer
dominate the population outside $2r_{200}$.  At the larger radii, the
CLEAN sample  dominates. The PREP
 sample accounts for  20 percent of the whole population outside
$2.5r_{200}$ and gradually drops to $5$ percent at $r_{200}$.  The
right panel shows results for more massive haloes in the mass range
$[10^{10},10^{11}]\ms$, where the respective fractions of $3$
populations are very similar to those of lower mass haloes as shown in
the left panel. Our results are consistent with those of \citet{Wang2009}
and  \citet{Bahe2012} who found a similar fraction of EJECT haloes
in their simulation, although for more massive haloes, suggesting that
the fraction of ejected haloes are largely independent of halo mass.

In following sections, we will only consider haloes with
cluster-centric radii of $[1.5, 3]r_{\rm 200}$.  We consider this
specific range for two reasons. 1) $3r_{200}$ is the radius inside
which the resolved dark haloes do not suffer from contamination from
heavy particles. 2) $1.5r_{200}$ is  roughly the scale of FOF groups corresponding to
Phoenix clusters. In these volumes around the nine Phoenix clusters, there are a total of
46000 haloes within mass range $[10^9, 10^{11}] \ms$, and the
fractions of the CLEAN, PREP, EJECT haloes are 45\%, 15\%, and 40\%,
respectively.

\subsection{Semi-analytical galaxies}
\label{sec:semi}
In order to assess  the effect of assembly bias  on
dwarf galaxy formation, we use the semi-analytical galaxy
formation model of  \citet[][hereafter G11]{Guo2011} to derive galaxy
properties.  With MS and MS-II simulations, the G11 SAM 
model well reproduces various observed galaxy properties. In
particular, it reproduces the stellar mass function of the Universe
across 5 orders of magnitude. In this work, we also apply the SAM
on the Phoenix simulations.

Galaxy formation and evolution is subjected to environmental effects，
which are reasonably modeled in the G11 SAM. For galaxies forming in the 
CLEAN haloes, the environmental dependence is expected to come from
the halo growth history. For example, in high density environments, 
the high velocity dispersion of dark matter particles and the local tidal fields
 may suppress the growth of low mass haloes
 \citep[e.g.][]{Wang2007, Dalal2008,  Hahn2009}, and suppress
the gas accretion in these haloes.  For the PREP and the EJECT halo samples,
additional effects come from interactions with their former host  haloes. When a
dark matter halo is accreted by a larger system,  tidal
forces strip both its baryonic and dark matter components. In
addition, ram pressure strips the hot gas component. Both mechanisms
suppress star formation and influence galaxy properties. Below, we
briefly describe the physical prescriptions of these two effects in the G11
model.

In the G11 model, the hot gas distribution in subhalo is assumed to parallel
that of the dark matter. Since the tidal force is identical for both components, 
the fraction of the stripped hot gas is in proportion to that of the dark
matter. The latter has already been followed in the original N-body
simulation. Thus, we have 
\begin{equation}
\frac{M_{\rm hot}(R_{\rm tidal})}{M_{\rm hot,infall}}= \frac{M_{\rm DM}}{M_{\rm DM,infall}},
\end{equation}
where $M_{\rm DM}$ and $M_{\rm hot}$ are the dark matter and the hot gas
masses of the subhalo,  while $M_{\rm DM,infall}$ and $M_{\rm
  hot,infall}$  are the masses of these two components at the time of infall.
If we assume that the hot gas follows a $r^{-2}$ profile,  the tidal radius
can be written as: %
\begin{equation}
R_{\rm tidal}=\left(\frac{M_{\rm DM}}{M_{\rm DM,infall}}\right)R_{\rm DM,infall},
\end{equation}
where $R_{\rm DM,infall}$ is the viral radius of the subhalo at its
infall. Beyond the tidal radius, the hot gas is assumed to be stripped
off. 

Unlike tidal stripping, ram pressure stripping only exerts on the hot
gas component. The gas pressure of a host halo at radius $R$ is $\rho_{\rm
hot}(R)V^2_{\rm orbit}$, where $\rho_{\rm hot}(R)$ is the hot gas density and
$V_{\rm orbit}$ is the orbit velocity at $R$.  The self-gravity of the subhalo
balances the host halo hot gas pressure at radii $R_{\rm r.p.}$, i.e.  
\begin{equation}
\rho_{\rm sat}(R_{\rm r.p.})V_{\rm sat}^2=\rho_{\rm par}(R)V_{\rm orbit}^2,
\end{equation}
where $\rho_{\rm sat}(R_{\rm r.p.})$ is the hot gas density of the
satellite at radius $R_{\rm r.p.}$, $V_{\rm sat}$ is the virial
velocity of the subhalo at infall (which is assumed to be constant as
the subhalo orbits around the main halo). The ram-pressure dominates
over gravity beyond $R_{\rm r.p.}$ and removes the hot gas of subhaloes
outside the radius. Therefore, the final stripping radius is defined as:

\begin{equation}
R_{\rm strip}=\min(R_{\rm tidal},R_{\rm r.p.}) \,.
\end{equation}
The stripped gas is added to the hot gas component of the host halo. In
the G11 model, stripping does not modify the gas profile within $R_{\rm strip}$.
Thus the cooling rate onto the center is not affected immediately. However,
removal of the hot gas suppresses gas cooling and quenches star formation
eventually. We refer readers to G11 for more details. 

\section{Halo properties}
\label{sec:haloresult}
In this section, we present the properties of dwarf-sized dark
matter haloes in the Phoenix simulations and compare them to those of
MS-II.
\subsection{Mass function}
\begin{figure}
\hspace{0.13cm}\resizebox{8cm}{!}{\includegraphics{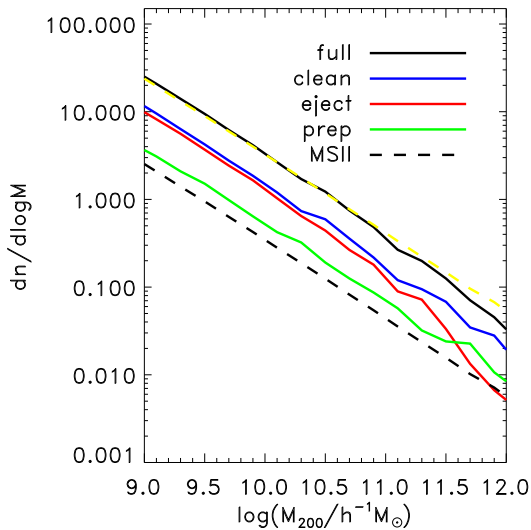}}
\caption{Halo mass functions. Black solid line: mean mass function
  in the regions around the  9 Phoenix clusters($[1.5, 3]r_{\rm 200}$);  blue line:
  CLEAN sample;  red line: EJECT sample; green line: PREP sample. The
  black dashed lines show the mass function of MS-II. In order to compare
  the shapes of halo mass function, we rescale the amplitude of the MS-II mass
  function vertically to match that of the full sample of the
  Phoenix(yellow dashed line).} \label{fig:mf}
\end{figure}

We first examine whether the shape of the halo mass function in the regions 
around the Phoenix clusters differs from the cosmic mean derived from the MS-II
simulation. The amplitude of the halo mass function is expected to be higher
in cluster regions because of the higher density, but it is not clear whether the
shape of the mass function changes with environment.  To this end, 
we show in Fig.~\ref{fig:mf} the halo mass function in the specified
volumes of the Phoenix and the MS-II simulations. The black solid line corresponds to 
the mass function of all surrounding haloes residing  within  $[1.5,
  3]r_{\rm 200}$ (full sample) of the $9$ Phoenix clusters, and the dashed
black line stands for the halo mass function of the MS-II simulation. For easier
comparison, we rescale the MS-II mass function vertically (yellow dashed
line) to match the amplitude of the full Phoenix sample. Clearly, the slopes of
the two halo mass functions are identical in all shown halo mass ranges
except for the high mass end, where the Poisson noise is largest. We also
show the mass functions of CLEAN, EJECT and PREP samples separately with
different lines. Again, there is no apparent difference at the low halo
mass end, while the mass function of EJECT sample is steeper above
halo mass  $10^{11}\ms$. This may reflect the fact that the massive
  haloes in the EJECT sample suffer strong stripping effect and lost
  substantial fraction of mass during interaction with massive
  clusters.

\subsection{Halo assembly history}

\begin{figure*}
\hspace{0.13cm}\resizebox{16cm}{!}{\includegraphics{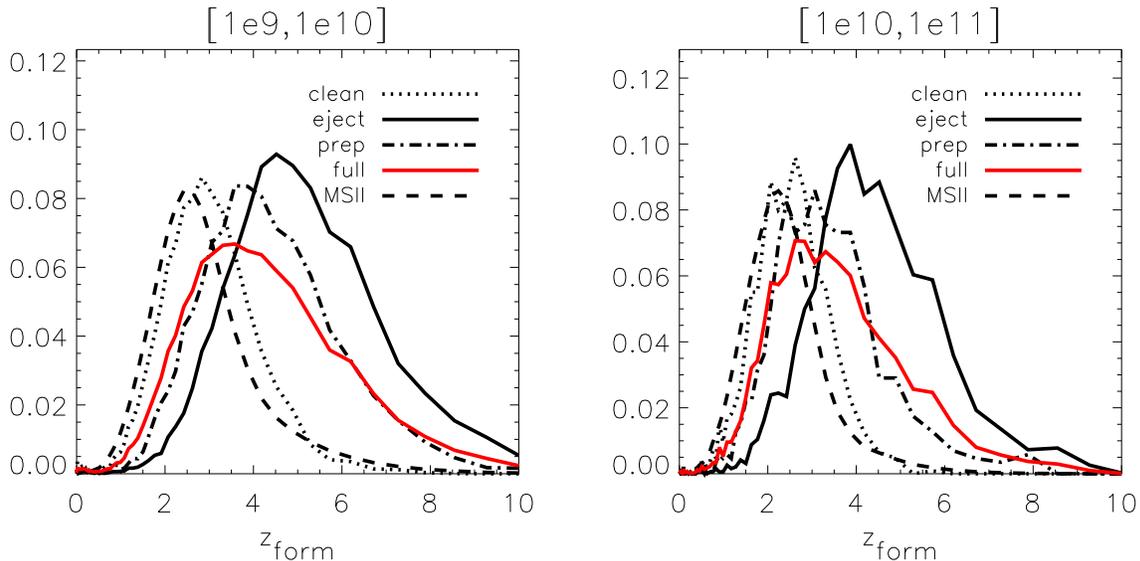}}
\caption{Formation time distribution of dark matter haloes. Dotted
  lines:  CLEAN sample; solid lines: EJECT sample; dash-dotted lines:
  PREP sample; dashed lines: MS-II. The left panel shows the result for
  $[10^9,10^{10}]\ms$ haloes while the right panel shows the result for
  $[10^{10},10^{11}]\ms$ haloes.}
\label{fig:zform}
\end{figure*}

We show the formation time distribution of dark haloes in different
simulations and for different halo samples in Fig.~\ref{fig:zform}. Here
we define the halo formation time to be the epoch at which the main
progenitor of the halo reaches the half of its present day mass. The results
 are shown for two  halo mass ranges: $[10^{9},10^{10}]\ms$ (left-hand panel)
and $[10^{10},10^{11}]\ms$ (right-hand panel). Compared to the
MS-II haloes,  the haloes in the regions around the Phoenix clusters form earlier. For  
haloes in the mass range of $[10^9,10^{10}]\ms$, the median formation time is about
$z\sim 4.5$ for the Phoenix haloes, while it is at a much later redshift $z \sim
2.5$ for the MS-II haloes. We also show the formation
time distribution for our three Phoenix halo samples. The
median formation redshift of the CLEAN sample, $z \sim
3$, is only slightly larger than MS-II average, while the PREP and
EJECT samples forms much earlier than the MS-II sample. In the right panel,
we show that the results for more massive haloes have the same trend.

\begin{figure*}
\hspace{0.13cm}\resizebox{16cm}{!}{\includegraphics{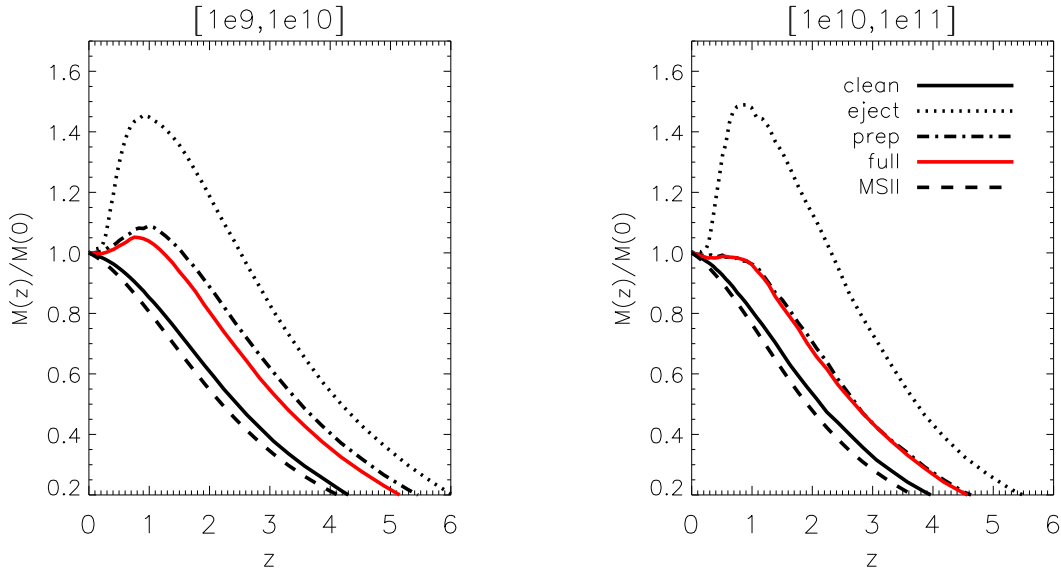}}
\caption{Halo mass growth history as a function of redshift. For each
  curve, median values are shown. Dotted lines: EJECT sample; solid
  lines: CLEAN sample; dash-dotted lines: PREP sample; dashed lines:
  MS-II haloes. Left panel shows results for haloes in mass range
  $[10^9,10^{10}]\ms$ while right panel shows the results for
  $[10^{10},10^{11}]\ms$ haloes.}
\label{fig:halogrowth}
\end{figure*}

We further explore the assembly history of the haloes in the
two mass ranges for different samples. In Fig.~\ref{fig:halogrowth},
we plot the growth history of haloes as a function of redshift. 
The growth history of the halo is defined as the ratio of its mass
at redshift $z$ to its present day mass, $M(z)/M(0)$. For different halo samples, 
the median values of $M(z)/M(0)$ are shown with different lines. 
For the CLEAN haloes, their mass increases monotonically
with time and their assembly is only slightly earlier than that of the MS-II haloes. 
The EJECT haloes acquire their mass much earlier than the other two samples. 
Statistically, EJECT haloes reach their maximum mass, which is $40$ percent
more than their present day value, at redshift $z \sim 1.2$, then lose a significant
fraction of mass afterwards due to their interactions with the clusters.  The same is true for
the PREP sample, except for the peak mass. On average, the Phoenix dwarf-sized
 haloes are $20$ percent more massive than the MS-II haloes at $z \sim1$,
 reflecting the fact that most interactions occur after redshift
2. The results for our two different mass ranges are similar.

\subsection{Halo structure}
\begin{figure}
\hspace{0.13cm}\resizebox{8cm}{!}{\includegraphics{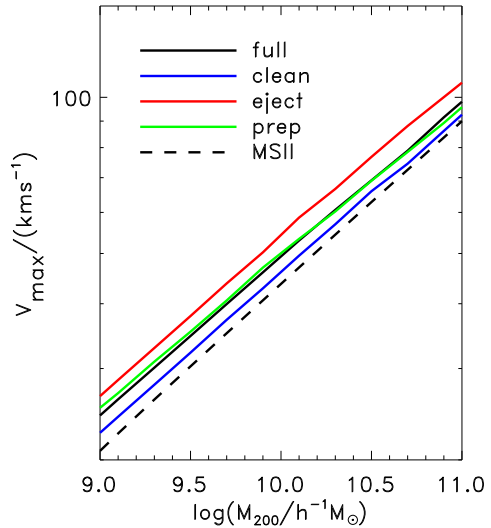}}
\caption{$V_{\rm max}$-$M_{\rm 200}$ relation for  dark matter haloes.
Median values are shown for different halo samples. Black solid:
  the full Phoenix sample; red: EJECT sample; green:  PREP sample; blue:
   CLEAN sample; black dashed lines:  MS-II haloes.}
\label{fig:vm}
\end{figure}

To examine how  assembly bias affects halo structure,  we plot the
maximum circular velocity of  dark matter haloes,  $V_{\rm max}$, as a
function of the halo mass for different halo samples in
Fig.~\ref{fig:vm}. For a given mass, the $V_{\rm max}$ of a halo is a proxy
for its concentration. More concentrated haloes have larger $V_{\rm
  max}$. For the dwarf-sized haloes of  the Phoenix simulations, the
 median $V_{\rm max}$ value is $20$ percent larger than that of the
 MS-II haloes. The differences are noticeable yet
very small between the CLEAN sample and the MS-II sample.  
The median  $V_{\rm max}$ values of PREP and  EJECT haloes are
$30\%$ and $25\%$ larger than that of MS-II haloes.  Recent studies
\citep[e.g.][]{Gao2012} show that while the tidal
force removes a significant fraction of dark matter from the subhalo,
it only reduces $V_{\rm max}$ slightly. Thus, the larger $V_{\rm
  max}$ in the Phoenix haloes may reflect the fact that these haloes were
more massive in the past and  interactions with larger haloes do not
reduce their maximum circular velocity significantly. On the other hand, the
small difference between the CLEAN sample and MS-II sample implies that 
the direct influence from large scale density field and large scale tidal field
on the structure of dwarf-sized haloes is weak.

  Note that, while the assembly bias effects shown above are
   mainly due to tidal interactions between the ejected haloes and their
   former hosts, it doesn't mean that the assembly bias seen in
   the clustering of haloes is also dominated by this
   mechanism. \citet{Dalal2008} argued that at the low mass end, the
   assembly bias in the clustering of haloes is due to the formation of a
   non-accreting population of haloes in the vicinity region of
   massive haloes. However, \citet{Keselman2007} found that assembly
   bias in the clustering of haloes is still present even after excluding
   the non-linear effects such as tidal disruption.  \citet{Wang2009}
   investigated the properties of ejected haloes in mass range from
   $10^{11}$ to $10^{12}\ms$, and found that though the ejected haloes
   have a much higher halo bias than the field haloes, they can not
   fully account for the assembly bias in the clustering of haloes
    because of their low abundance (less than 10\%).  The large scale
    environmental effects such as large scale tidal field also play an
    important role in assembly bias in the clustering of
    haloes\citep[e.g.][]{Wang2007,Hahn2007,Hahn2009}.

\section{Galaxy properties}
\begin{figure}
\includegraphics[width=0.4\textwidth]{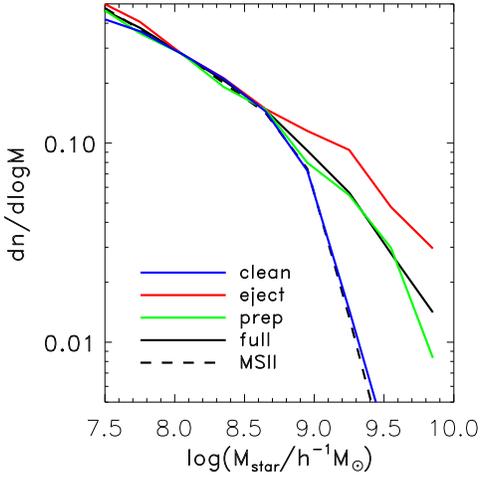}
\caption{The stellar mass function for  galaxies whose host haloes are in the
 mass range  $[10^9,10^{11}]\ms$. Solid lines: Phoenix galaxies; dashed line:
  MS-II galaxies. The MS-II stellar mass function is scaled to match the
  amplitude of the Phoenix stellar mass function at $10^8\ms$.}
\label{fig:smf}
\end{figure}
\begin{figure}
\hspace{0.13cm}\resizebox{8cm}{!}{\includegraphics{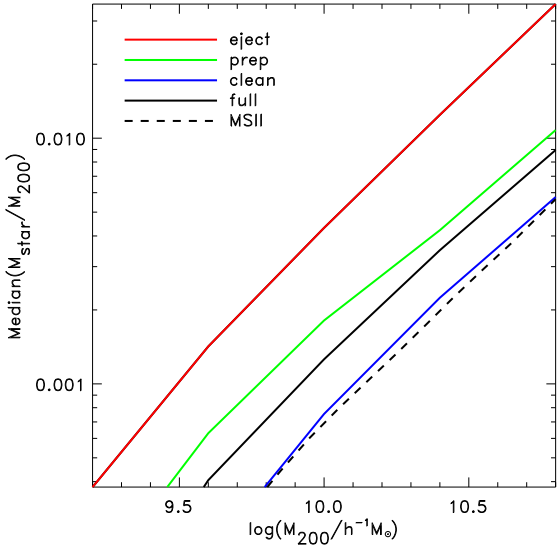}}
\caption{The median stellar mass-halo mass ratio as a function of stellar mass for
 EJECT(red),  PREP(green) and CLEAN(blue) galaxies. The black solid line shows
the result of  the full Phoenix dwarf haloes. The MS-II result is
shown with a black dashed line. }
\label{fig:gala_ej_ML}
\end{figure}

\label{sec:galaxyresult}

As shown in the previous sections,   assembly bias affects the properties
of dark matter haloes significantly. It may in turn influence galaxy properties
 such as halo occupation, color, etc. In this work, We explore the effects of assembly
 bias on the properties of dwarf galaxies in the  
 galaxy formation model of \citet{Guo2011},  which incorporates 
 all the necessary baryon physics.

We divide the Phoenix galaxies into different categories according to
their host halo samples.  We refer to the galaxies in the EJECT, PREP
and CLEAN haloes as the EJECT, PREP and CLEAN galaxies, respectively.
Here we also use the MS-II galaxy catalog of G11 to represent the galaxy population in the field.
 For a fair and easy comparison, we only 
consider  galaxies which reside in the haloes considered in
previous sections, i.e. within the mass range of $[10^{9},10^{11}]$ $\ms$.

Fig.~\ref{fig:smf} shows the stellar mass functions of galaxies. 
 In order to compare the shapes of the stellar mass functions in different halo samples, 
 we normalize these curves at the same stellar mass scale ${1 \times 10^8}\ms$. Clearly the
stellar mass functions of different samples have identical shapes at the low mass
end but start to differ at the stellar mass of $\log(M_{\rm star}) \sim
8.6$.  At the high mass end, the slope of the stellar mass function is
flatter for the galaxies in the regions around clusters. The
difference becomes larger as mass increases. For Phoenix galaxies, 
the stellar mass function of EJECT galaxies has very flat slope at the high mass end, and the
mass function of CLEAN galaxies roughly follows
that of MS-II. The difference of the 
stellar mass functions at the high mass end between 
    Phoenix and MS-II galaxies is caused by our restriction on the host
    halo mass range. Since tidal striping causes significant mass loss in EJECT and PREP haloes,
  galaxies near clusters occupy less massive haloes than the field galaxies
  of the same stellar mass (also see Fig.~\ref{fig:gala_ej_ML}). Therefore, for a given upper mass 
  limit of haloes, the fraction of massive galaxies is larger for the Pheonix sample.
  Note that, without the restriction on the halo mass, the shape of stellar mass
  function of the two simulations will be very similar, in agreement with observation results of
   \citet{Calvi2013}.

We show in Fig.~\ref{fig:gala_ej_ML} the ratio of stellar mass to halo
mass as a function of halo mass for the Phoenix (black solid) and MS-II
simulations (black dashed). The ratios for both simulations have very
similar power law relations with halo mass, while the amplitude of the
former is a factor of 2 higher. The results for our 3 samples of Phoenix
galaxies are also shown in the same plot. For a given halo
mass, the median stellar mass fraction of the EJECT halo sample is a
factor of 5 higher than that of the MS-II halo sample. This can be easily 
understood within the SAM framework that the stellar component of a dwarf
galaxy changes little even when a large fraction of  the dark matter is stripped. For
the CLEAN sample, the stellar mass to halo mass ratio is only about 10 percent
higher than that of the MS-II, suggesting that the environmental dependence
of the dwarf galaxies mainly comes from the tidal interactions between the ejected haloes
and their former host haloes. Recent work by  \citet{Wetzel2013} explored
the stellar mass-halo mass ratio using the abundance matching method. For
galaxies in the $[10^{9},10^{11}\ms]$ stellar mass range, they find this ratio
 for the ejected galaxies is $2.5$ times higher than that of field galaxies. This is
  similar to  the results found  in our work for lower mass
  galaxies. We expect that this environmental dependence can be
  measured with future galaxy-galaxy lensing surveys \citep{Li2013}.

\begin{figure}
\hspace{0.13cm}\resizebox{8cm}{!}{\includegraphics{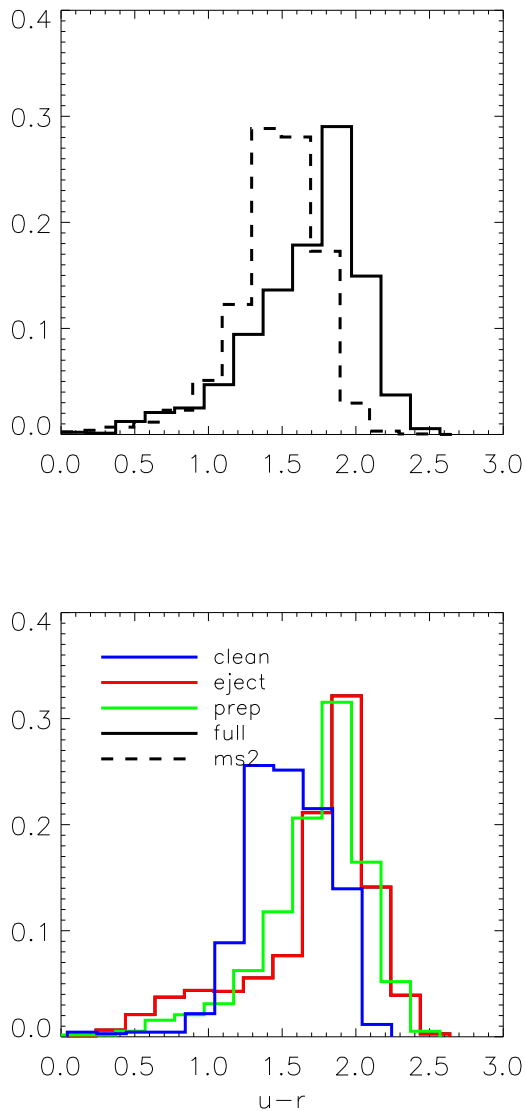}}
\caption{Histogram of u-r color  for the Phoenix and  MS-II
  galaxies.  Black solid histogram shows color distribution for 
  the dwarf galaxies in the Phoenix simulations with cluster-centric radii
  of $[1.5 ,3]r_{\rm 200}$ and  the dashed histogram shows  galaxies in the 
  MS-II run. The u-r colors for the CLEAN, EJECT and PREP galaxies are shown as blue, 
   red and green solid lines, respectively.}
\label{fig:color}
\end{figure}

As discussed before,  interactions with massive haloes remove the hot
gas and therefore suppress star formation activity, which in
turn may affect the color of galaxies. In Fig.~\ref{fig:color}, we compare
the SDSS u-r color for the Phoenix and MS-II galaxies. Dwarf galaxies
near Phoenix clusters are on average $\sim 0.5$ magnitudes redder than 
MS-II galaxies, while the color distribution of the CLEAN galaxies is almost
the same with that of MS-II galaxies.   We also show the colors for our three
Phoenix galaxy samples separately. The EJECT and PREP samples are much
redder than the CLEAN sample. Combining the results shown in
Fig.\ref{fig:color} and Fig.\ref{fig:gala_ej_ML}, we expect an
enhanced stellar-halo mass ratio to be seen in the reddest dwarf
galaxies near clusters.

Recent studies show that there might be tension between the
  G11 model and observational data. It was shown in  \citet{Guo2011}
  and \citet{Weinmann2012} that the low mass galaxies in G11 model are
  too red. \citet{Weinmann2011} also found that the galaxies in G11
  model seem to form too early. In addition, \citet{Weinmann2012}
  showed that though G11 model well reproduces the red galaxy fraction
  in clusters for Coma and Perseus, it fails to explain the low red
  fraction of faint galaxies in Virgo cluster, which may be partly due
  to the treatment of environmental effects in G11 model.
  \citet{Wang2013} studied the assembly bias by investigating the
  dependence of the clustering of central galaxies on the specific
  star formation rate and found that the  observed assembly bias in
  SDSS data is weaker than that predicted by the G11 SAM. These studies
  suggest that assembly bias effects may be over estimated by the G11
  model. Therefore, it is important to confront our predictions with
  real observational data in future.

\section{Discussion and summary}
\label{sec:sum}
In this paper, we use the state-of-art dark matter simulations to
explore how  assembly bias affects the properties of  
dwarf-sized dark matter haloes and their  galaxies.  We divide our haloes into three
 different samples according to their assembly history and investigate 
 the origin of assembly bias effects. We then combine
 our simulations with the semi-analytical galaxy model of G11 to examine
 how these effects influence the observable properties of galaxies.  
 Our main results can be summarized as follows:

(i) Among all dwarf-sized haloes distributed in a cluster-centric range of
$[1.5,3]r_{200}$ around rich cluster haloes, about $40\%$ have been
accreted by the cluster haloes in the past(EJECT haloes).  The fraction increases
with cluster-centric radii decreasing. About $45\%$  of haloes in the same region 
have ordinary growth tracks, i.e. have no substantial interactions with
larger objects (CLEAN sample).  The remaining $15\%$ of the haloes have been
accreted by haloes other than the cluster haloes in the past
(PREP sample).  These respective fractions are largely independent of
halo mass.

(ii) Dwarf-sized haloes near clusters on average forms earlier than
the haloes in the field by a redshift difference $\delta z \sim 2$.
The maximum circular velocity of dwarf haloes near the clusters is
also 20\% larger. 

(iii) On average, the mass of the EJECT halo sample reaches a peak at 
about redshift 1,  at $\sim 1.4$ times the present day mass. In
contrast, the mass of CLEAN haloes grow monotonically
 with redshift. 

(iv) For a given halo mass, dwarf-sized haloes near clusters  have 
three times as much stellar mass compared to the field haloes. This factor is
largely independent of halo mass. Such a difference may be confirmed
by future observations, such as galaxy-galaxy lensing surveys \citep{Li2013}. 

(v) Dwarf galaxies in the regions around clusters are 0.5 magnitude redder
 than the galaxies in the field. 
 
 When considering the different origins of assembly bias, we find that
the differences seen in halo and galaxy properties are largely
due to tidal interactions between the ejected haloes and their former
host haloes, while the influences from the large scale environment
such as the tidal field and the velocity field are weak. It should
be possible to explore  the assembly bias effects on dwarf galaxies in 
observations.

\section*{Acknowledgements}
Some of simulations used in this work were carried out on the Lenova
Deepcomp7000 supercomputer of the super Computing Centre of Chinese
Academy of Sciences, Beijing, China. LR is supported by China Postdoctoral
Science Foundation, Grant NO. 2011M500395 and NAOC Science Funds for Youth 
Scholar. LG acknowledges support from the
one-hundred-talents program of the Chinese academy of science (CAS),
the National Basic Research Program of China (program 973 under grant
No. 2009CB24901), {\small NSFC} grants (Nos. 10973018 and 11133003),
{\small MPG} partner Group family, and an {\small STFC} Advanced
Fellowship, as well as the hospitality of the Institute for
Computational Cosmology at Durham University. The authors 
acknowledge the suggestions of Dr. Lan Wang and Dr. Andrew Cooper.

\label{lastpage}
\bibliographystyle{mnras}
\bibliography{ref}

\begin{thebibliography}{52}
\expandafter\ifx\csname natexlab\endcsname\relax\def\natexlab#1{#1}\fi

\bibitem[{Bah{\'e}} et~al.(2012){Bah{\'e}}, {McCarthy}, {Crain} \&
  {Theuns}]{Bahe2012}
{Bah{\'e}} Y.~M., {McCarthy} I.~G., {Crain} R.~A., {Theuns} T., 2012, \mnras,
  424, 1179

\bibitem[{Balogh} et~al.(2000){Balogh}, {Navarro} \& {Morris}]{Balogh2000}
{Balogh} M.~L., {Navarro} J.~F., {Morris} S.~L., 2000, \apj, 540, 113

\bibitem[{Benson}(2005)]{Benson2005}
{Benson} A.~J., 2005, \mnras, 358, 551

\bibitem[{Berrier} et~al.(2009){Berrier}, {Stewart}, {Bullock}, {Purcell},
  {Barton} \& {Wechsler}]{Berrier2009}
{Berrier} J.~C., {Stewart} K.~R., {Bullock} J.~S., {Purcell} C.~W., {Barton}
  E.~J., {Wechsler} R.~H., 2009, \apj, 690, 1292

\bibitem[{Blanton} \& {Berlind}(2007)]{Blanton2007}
{Blanton} M.~R., {Berlind} A.~A., 2007, \apj, 664, 791

\bibitem[{Boylan-Kolchin} et~al.(2009){Boylan-Kolchin}, {Springel}, {White},
  {Jenkins} \& {Lemson}]{BK2009}
{Boylan-Kolchin} M., {Springel} V., {White} S.~D.~M., {Jenkins} A., {Lemson}
  G., 2009, \mnras, 398, 1150

\bibitem[{Calvi} et~al.(2013){Calvi}, {Poggianti}, {Vulcani} \&
  {Fasano}]{Calvi2013}
{Calvi} R., {Poggianti} B.~M., {Vulcani} B., {Fasano} G., 2013, \mnras, 432,
  3141

\bibitem[{Colless} et~al.(2001){Colless}, {Dalton}, {Maddox}
  et~al.]{Colless2001}
{Colless} M., {Dalton} G., {Maddox} S., et~al., 2001, \mnras, 328, 1039

\bibitem[{Cooper} et~al.(2010){Cooper}, {Gallazzi}, {Newman} \&
  {Yan}]{Cooper2010}
{Cooper} M.~C., {Gallazzi} A., {Newman} J.~A., {Yan} R., 2010, \mnras, 402,
  1942

\bibitem[{Dalal} et~al.(2008){Dalal}, {White}, {Bond} \& {Shirokov}]{Dalal2008}
{Dalal} N., {White} M., {Bond} J.~R., {Shirokov} A., 2008, \apj, 687, 12

\bibitem[{Davis} et~al.(1985){Davis}, {Efstathiou}, {Frenk} \&
  {White}]{Davis1985}
{Davis} M., {Efstathiou} G., {Frenk} C.~S., {White} S.~D.~M., 1985, \apj, 292,
  371

\bibitem[{Desjacques}(2008)]{Desjacques2008}
{Desjacques} V., 2008, \mnras, 388, 638

\bibitem[{Duffy} et~al.(2008){Duffy}, {Schaye}, {Kay} \& {Dalla
  Vecchia}]{Duffy2008}
{Duffy} A.~R., {Schaye} J., {Kay} S.~T., {Dalla Vecchia} C., 2008, \mnras, 390,
  L64

\bibitem[{Gao} et~al.(2012){Gao}, {Navarro}, {Frenk}, {Jenkins}, {Springel} \&
  {White}]{Gao2012}
{Gao} L., {Navarro} J.~F., {Frenk} C.~S., {Jenkins} A., {Springel} V., {White}
  S.~D.~M., 2012, \mnras, 425, 2169

\bibitem[{Gao} et~al.(2005){Gao}, {Springel} \& {White}]{Gao2005}
{Gao} L., {Springel} V., {White} S.~D.~M., 2005, \mnras, 363, L66

\bibitem[{Gao} \& {White}(2007)]{Gao2007}
{Gao} L., {White} S.~D.~M., 2007, \mnras, 377, L5

\bibitem[{Guo} et~al.(2011){Guo}, {White}, {Boylan-Kolchin} et~al.]{Guo2011}
{Guo} Q., {White} S., {Boylan-Kolchin} M., et~al., 2011, \mnras, 413, 101

\bibitem[{Hahn} et~al.(2007){Hahn}, {Carollo}, {Porciani} \& {Dekel}]{Hahn2007}
{Hahn} O., {Carollo} C.~M., {Porciani} C., {Dekel} A., 2007, \mnras, 381, 41

\bibitem[{Hahn} et~al.(2009){Hahn}, {Porciani}, {Dekel} \& {Carollo}]{Hahn2009}
{Hahn} O., {Porciani} C., {Dekel} A., {Carollo} C.~M., 2009, \mnras, 398, 1742

\bibitem[{Harker} et~al.(2006){Harker}, {Cole}, {Helly}, {Frenk} \&
  {Jenkins}]{Harker2006}
{Harker} G., {Cole} S., {Helly} J., {Frenk} C., {Jenkins} A., 2006, \mnras,
  367, 1039

\bibitem[{Hinshaw} et~al.(2012){Hinshaw}, {Larson}, {Komatsu} et~al.]{Wmap2012}
{Hinshaw} G., {Larson} D., {Komatsu} E., et~al., 2012, ArXiv:astro-ph/1212.5226

\bibitem[{Jing} et~al.(1998){Jing}, {Mo} \& {Boerner}]{Jing1998}
{Jing} Y.~P., {Mo} H.~J., {Boerner} G., 1998, \apj, 494, 1

\bibitem[{Jing} et~al.(2007){Jing}, {Suto} \& {Mo}]{Jing2007}
{Jing} Y.~P., {Suto} Y., {Mo} H.~J., 2007, \apj, 657, 664

\bibitem[{Kauffmann} et~al.(2013){Kauffmann}, {Li}, {Zhang} \&
  {Weinmann}]{Kauffmann2013}
{Kauffmann} G., {Li} C., {Zhang} W., {Weinmann} S., 2013, \mnras, 430, 1447

\bibitem[{Keselman} \& {Nusser}(2007)]{Keselman2007}
{Keselman} J.~A., {Nusser} A., 2007, \mnras, 382, 1853

\bibitem[{Lacerna} \& {Padilla}(2011)]{Lacerna2011}
{Lacerna} I., {Padilla} N., 2011, \mnras, 412, 1283

\bibitem[{Lacerna} \& {Padilla}(2012)]{Lacerna2012}
{Lacerna} I., {Padilla} N., 2012, \mnras, 426, L26

\bibitem[{Li} et~al.(2013){Li}, {Mo}, {Fan}, {Yang} \& {Bosch}]{Li2013}
{Li} R., {Mo} H.~J., {Fan} Z., {Yang} X., {Bosch} F.~C.~v.~d., 2013, \mnras,
  430, 3359

\bibitem[{Li} et~al.(2008){Li}, {Mo} \& {Gao}]{Li2008}
{Li} Y., {Mo} H.~J., {Gao} L., 2008, \mnras, 389, 1419

\bibitem[{Ludlow} et~al.(2009){Ludlow}, {Navarro}, {Springel}, {Jenkins},
  {Frenk} \& {Helmi}]{Ludlow2009}
{Ludlow} A.~D., {Navarro} J.~F., {Springel} V., {Jenkins} A., {Frenk} C.~S.,
  {Helmi} A., 2009, \apj, 692, 931

\bibitem[{McGee} et~al.(2009){McGee}, {Balogh}, {Bower}, {Font} \&
  {McCarthy}]{McGee2009}
{McGee} S.~L., {Balogh} M.~L., {Bower} R.~G., {Font} A.~S., {McCarthy} I.~G.,
  2009, \mnras, 400, 937

\bibitem[{Peacock} \& {Smith}(2000)]{Peacock2000}
{Peacock} J.~A., {Smith} R.~E., 2000, \mnras, 318, 1144

\bibitem[{Planck Collaboration} et~al.(2013){Planck Collaboration}, {Ade},
  {Aghanim} et~al.]{Planck2013}
{Planck Collaboration}, {Ade} P.~A.~R., {Aghanim} N., et~al., 2013, ArXiv: astroph/1303.5076
  

\bibitem[{Sales} et~al.(2007){Sales}, {Navarro}, {Abadi} \&
  {Steinmetz}]{Sales2007}
{Sales} L.~V., {Navarro} J.~F., {Abadi} M.~G., {Steinmetz} M., 2007, \mnras,
  379, 1475

\bibitem[{Sandvik} et~al.(2007){Sandvik}, {M{\"o}ller}, {Lee} \&
  {White}]{Sandvik2007}
{Sandvik} H.~B., {M{\"o}ller} O., {Lee} J., {White} S.~D.~M., 2007, \mnras,
  377, 234

\bibitem[{Skibba} \& {Sheth}(2009)]{Skibba2009}
{Skibba} R.~A., {Sheth} R.~K., 2009, \mnras, 392, 1080

\bibitem[{Spergel} et~al.(2003){Spergel}, {Verde}, {Peiris}
  et~al.]{spergel2003}
{Spergel} D.~N., {Verde} L., {Peiris} H.~V., et~al., 2003, \apjs, 148, 175

\bibitem[{Springel} et~al.(2005){Springel}, {White}, {Jenkins}
  et~al.]{Springel2005}
{Springel} V., {White} S.~D.~M., {Jenkins} A., et~al., 2005, \nat, 435, 629

\bibitem[{Springel} et~al.(2001){Springel}, {Yoshida} \&
  {White}]{Springel2001b}
{Springel} V., {Yoshida} N., {White} S.~D.~M., 2001, Nature, 6, 79

\bibitem[{Tinker} et~al.(2011){Tinker}, {Wetzel} \& {Conroy}]{Tinker2011}
{Tinker} J., {Wetzel} A., {Conroy} C., 2011, ArXiv:astroph/1107.5046

\bibitem[{Tinker} et~al.(2008){Tinker}, {Conroy}, {Norberg}, {Patiri},
  {Weinberg} \& {Warren}]{Tinker2008}
{Tinker} J.~L., {Conroy} C., {Norberg} P., {Patiri} S.~G., {Weinberg} D.~H.,
  {Warren} M.~S., 2008, \apj, 686, 53

\bibitem[{Tinker} et~al.(2012){Tinker}, {George}, {Leauthaud}
  et~al.]{Tinker2012}
{Tinker} J.~L., {George} M.~R., {Leauthaud} A., et~al., 2012, \apjl, 755, L5

\bibitem[{Wang} et~al.(2009){Wang}, {Mo} \& {Jing}]{Wang2009}
{Wang} H., {Mo} H.~J., {Jing} Y.~P., 2009, \mnras, 396, 2249

\bibitem[{Wang} et~al.(2007){Wang}, {Mo} \& {Jing}]{Wang2007}
{Wang} H.~Y., {Mo} H.~J., {Jing} Y.~P., 2007, \mnras, 375, 633

\bibitem[{Wang} et~al.(2013){Wang}, {Weinmann}, {De Lucia} \& {Yang}]{Wang2013}
{Wang} L., {Weinmann} S.~M., {De Lucia} G., {Yang} X., 2013, \mnras

\bibitem[{Wang} et~al.(2008){Wang}, {Yang}, {Mo}, {van den Bosch}, {Weinmann}
  \& {Chu}]{Wang2008}
{Wang} Y., {Yang} X., {Mo} H.~J., {van den Bosch} F.~C., {Weinmann} S.~M.,
  {Chu} Y., 2008, \apj, 687, 919

\bibitem[{Wechsler} et~al.(2006){Wechsler}, {Zentner}, {Bullock}, {Kravtsov} \&
  {Allgood}]{Wechsler2006}
{Wechsler} R.~H., {Zentner} A.~R., {Bullock} J.~S., {Kravtsov} A.~V., {Allgood}
  B., 2006, \apj, 652, 71

\bibitem[{Weinmann} et~al.(2011){Weinmann}, {Lisker}, {Guo}, {Meyer} \&
  {Janz}]{Weinmann2011}
{Weinmann} S.~M., {Lisker} T., {Guo} Q., {Meyer} H.~T., {Janz} J., 2011,
  \mnras, 416, 1197

\bibitem[{Weinmann} et~al.(2012){Weinmann}, {Pasquali}, {Oppenheimer}
  et~al.]{Weinmann2012}
{Weinmann} S.~M., {Pasquali} A., {Oppenheimer} B.~D., et~al., 2012, \mnras,
  426, 2797

\bibitem[{Wetzel} et~al.(2013){Wetzel}, {Tinker}, {Conroy} \& {van den
  Bosch}]{Wetzel2013}
{Wetzel} A.~R., {Tinker} J.~L., {Conroy} C., {van den Bosch} F.~C., 2013,
  \mnras, 432, 336

\bibitem[{Yang} et~al.(2003){Yang}, {Mo} \& {van den Bosch}]{Yang2003}
{Yang} X., {Mo} H.~J., {van den Bosch} F.~C., 2003, \mnras, 339, 1057

\bibitem[{Yang} et~al.(2006){Yang}, {Mo} \& {van den Bosch}]{Yang2006}
{Yang} X., {Mo} H.~J., {van den Bosch} F.~C., 2006, \apjl, 638, L55

\end{thebibliography}

\end{document}